\documentclass[iop]{emulateapj}

\usepackage{longtable}
\usepackage{afterpage}

\def\kms{$\textrm{km s}^{-1}$}

\begin{document}

\shorttitle{An intermediate-scale bipole observed with IMaX}
\shortauthors{Guglielmino et al.}

\title{The frontier between Small-scale bipoles and Ephemeral Regions in the solar photosphere: Emergence and Decay of an Intermediate-scale bipole\\observed with IMaX/SUNRISE}

\author{S.~L. Guglielmino\altaffilmark{1,2}, V. Mart\'inez Pillet\altaffilmark{1},
J.~A. Bonet\altaffilmark{1,2}, J.~Carlos del Toro Iniesta\altaffilmark{3}, L.~R. Bellot Rubio\altaffilmark{3},
S.~K. Solanki\altaffilmark{4,5}, W. Schmidt\altaffilmark{6}, A. Gandorfer\altaffilmark{4}, P. Barthol\altaffilmark{4}, M. Kn\"olker\altaffilmark{7}}
\altaffiltext{1}{IAC Instituto de Astrof\'isica de Canarias, C/ V\'ia L\'actea s/n, La Laguna, Tenerife, E-38200, Spain}
\altaffiltext{2}{ULL Departamento de Astrof\'isica, Univ. de La Laguna, La Laguna, Tenerife, E-38205, Spain}
\altaffiltext{3}{IAA Instituto de Astrof\'isica de Andaluc\'ia (CSIC), Apdo.\ de Correos 3004, E 18080 Granada, Spain}
\altaffiltext{4}{MPS Max-Planck-Institut f\"ur Sonnensystemforschung, Max-Planck-Str. 2, 37191 Katlenburg-Lindau, Germany}
\altaffiltext{5}{School of Space Research, Kyung Hee University, Yongin, Gyeonggi 446-701, Korea}
\altaffiltext{6}{KIS Kiepenheuer-Institut f\"ur Sonnenphysik, Sch\"oneckstr. 6, 79104 Freiburg, Germany}
\altaffiltext{7}{HAO High Altitude Observatory, National Center for Atmospheric Research, P.O. Box 3000, Boulder, CO 80307-3000, USA}

\email{sgu@iac.es}

\begin{abstract}

We report on the photospheric evolution of an intermediate-scale ($\approx 4$
Mm footpoint separation) magnetic bipole, from emergence to decay, observed in
the quiet Sun at high spatial (0\farcs3) and temporal (33 s) resolution. The
observations were acquired by the IMaX imaging magnetograph during the first
science flight of the \textsc{Sunrise} balloon-borne solar observatory. The
bipole flux content is $6 \times 10^{17}$ Mx, representing a structure bridging
the gap between granular scale bipoles and the smaller ephemeral regions.
Footpoints separate at a speed of 3.5 \kms{} and reach a maximum distance of
4.5 Mm before the field dissolves. The evolution of the bipole is revealed to
be very dynamic: we found a proper motion of the bipole axis and detected a
change of the azimuth angle of 90$^{\circ}$ in 300 seconds. The overall
morphology and behaviour are in agreement with previous analyses of bipolar
structures emerging at granular scale, but we also found several similarities
with emerging flux structures at larger scale. The flux growth rate is $2.6
\times 10^{15}$ Mx s$^{-1}$, while the mean decay rate is one order of
magnitude smaller. 
We describe in some detail the decay phase of the bipole footpoints which
includes break up into smaller structures, interaction with pre-existing fields
leading to cancellation but appears to be dominated by an as-yet unidentified
diffusive process that removes most of the flux with an exponential flux decay
curve. The diffusion constant ($8 \times 10^{2} \;\textrm{km}^{2}~\textrm{s}^{-1}$) 
associated with this decay is similar to the values used to describe the large scale 
diffusion in flux transport models.

\end{abstract}


\section{Introduction}

Magnetic flux emergence in the solar photosphere involves a variety of spatial
scales. It spans from the small bipoles with fluxes of the order of $10^{16}$
Mx \citep{Marian:09,Danilovic:10}, that populate the quiet Sun at any stage of 
the solar activity cycle, to the large, complex active regions present during 
solar maximum, with fluxes of the order of $10^{22}$ Mx \citep{Lites:98}. 

Bipolar structures with absolute flux content ranging from $10^{18}$ Mx to $5
\times 10^{19}$ Mx are usually referred to as ephemeral regions (ERs), due to 
the short lifetime of less than 24 hours that was estimated when they were first
discovered \citep{HarveyMartin:73}. The opposite polarity components of ERs
separate from each other during the growth period,
reaching a distance of a few megameters in about half an hour with decreasing
separation velocity \citep{Martin:88}. Then, ERs decay in a complex way,
strongly dependent upon the surrounding magnetic network elements. 

This evolution scenario was confirmed by the study of \citet{Hag:01}, who analyzed a
large statistical sample of ERs using time series of magnetograms taken with
the Michelson Doppler Imager (MDI) aboard the \emph{SOHO} satellite. She found an
average lifetime of $\simeq 4$ hours, an average separation of $\simeq 9$ Mm,
depending on the total amount of magnetic flux carried by the ER, and a separation velocity
decreasing from 4 \kms{} to 1.5 \kms. The flux concentrations fragmented and
moved apart, merging with other concentrations or with intranetwork fields. 

The origin of ERs is quite controversial. They could represent the small-scale
tail of the distribution of active regions, generated by the large
scale solar dynamo \citep{Parker:55}, as argued by \citet{Harvey:75}, or be
generated by a near-surface local dynamo \citep{Nordlund:92,Parker:93,Cattaneo:99}. 
Alternatively, they might be the end product of failed active region emergence processes 
that end in a catastrophic explosion \citep{MorenoInsertis:95}.

To shed light into their origin, \citet{Hag:01} also analyzed the relationship of 
ERs with the solar magnetic activity, concluding that only 60\% of the ERs had an orientation 
consistent with Hale's polarity law. \citet{Hag:03} found that the number
of ERs varies by a factor of 1.5 in antiphase with the solar cycle, confirming the
early findings of \citet{MartinHarvey:79}, and that the latitude distribution
of ERs is broader than that of the activity belts. The
frequency spectrum of ERs deduced by \citet{Hag:03} appeared to be a
continuous, smoothly decreasing distribution of bipolar flux regions, formed by
two distinct but coupled power-law spectra, with a turnover at around $2 \times
10^{19}$ Mx. This suggested the coexistence of both dynamo mechanisms, the
global dynamo producing cycle-modulated large ERs and active regions, and the
local dynamo generating small-scale bipoles independent of the cycle phase.
Furthermore, \citet{Hag:08} showed that the rate of ERs emergence depends on
the local flux imbalance: it is lower within strongly unipolar regions by a
factor of 3 relative to flux-balanced areas in the quiet Sun. This indicated that
a large background field may affect the ERs generation process in the
sub-surface layers.

In the meantime, our knowledge of small-scale magnetic fields greatly improved
from below the ERs threshold at $10^{18}$ Mx down to $10^{16}$ Mx, at the limit
of current detection capabilities. Thanks to the high-resolution
observations taken by both ground-based telescopes with adaptive optics,
and the Solar Optical Telescope
\citep[SOT;][]{Tsuneta:08} aboard the \emph{Hinode} satellite
\citep{Kosugi:07}, a number of small low-flux bipolar
structures have been studied in recent years. 
Several contributions reported on the emergence of horizontally inclined
magnetic fields with strengths of a few hundred G in the quiet Sun, mostly at
granular scale \citep{Marian:07,Centeno:07,Orozco:08,Lites:08,Ishikawa:08,
Marian:09,Gomory:10,Ishikawa:10,Danilovic:11}. The usual evolution of these structures
begins with the appearance of an isolated patch of linear polarization signals,
indicating horizontal fields, followed by the appearance of opposite-polarity
circular polarization signals at opposite edges of the horizontal patch. These
results are often interpreted as the rise of small-scale magnetic loops through
the photosphere. 

Different mechanisms have been proposed to explain the decay
of the bipoles: submergence of $\Omega$-loops driven by convective downflows, 
flux dispersal resulting from horizontal convective motions with the Stokes signals
falling below the instrumental sensitivity, ohmic dissipation, and the rise of U-loops
towards the upper photospheric layers following reconnection below the solar surface. 

\citet{Parnell:09} studied MDI and SOT data to determine the flux distribution of all 
magnetic features that can be observed with these instruments on the solar surface at
any one time. They found that flux concentrations can be described by
a single power-law distribution with a slope of -1.85
for fluxes between $2 \times 10^{17}$ Mx to $10^{23}$ Mx. These authors argue that the
differences in flux distributions found in previous works result from
counting and identifying the magnetic structures in different ways, i.e., from
selection effects. Their findings support a scale invariance of the
distribution of fluxes that conceivably may extend to even smaller magnetic features
beyond the current detection limit. Such a distribution might be produced
either (i) by a solar dynamo that operates in the same way at a continuum of
scales, and whose length scale increases with depth or (ii) if the magnetic
features are generated by distinct dynamos with different characteristic lengths,
by reprocessing of flux at the surface. In the latter case, such a process,
through merging, cancellation, and fragmentation, would dominate to produce a power-law distribution. 

\citet{Thornton:11} investigated the distribution of small-scale intranetwork
concentrations counting only the features \emph{related to emergence events} with
fluxes smaller than $10^{17}$ Mx. By combining their results with the results on
the distributions of the emergence at larger scale, they found a single
power law for all emerging features with fluxes ranging from $10^{16}$ Mx to $10^{23}$ Mx 
with a slope of -2.7, which is consistent with the findings of
\citet{Hag:03}. The emergence rate is dominated by small-scale fields, in
agreement with \citet{Socas:02}. Since the single power-law distribution is
already found at emergence, before the surface flux reprocessing, this
result is suggestive of the idea that dynamo action occurs on all
scales. Large magnetic structures would be created in the tachocline, whilst
the smaller features would continuously occur by a turbulent dynamo action over
a range of scales throughout the convection zone. Numerical convection
simulations also support this scenario \citep{Stein:10}. The steeper slope of
emerging structures (-2.7) with respect to the magnetic features present at any
instant on the solar surface (-1.85) can be explained taking into account the shorter
lifetime of the small-scale fields and their coalescence at the edges of the
supergranules. 

In this paper, we report on the emergence of a magnetic bipole, observed at
high spatial resolution by the Imaging Magnetograph eXperiment
\citep[IMaX;][]{Valentin:11} mounted on the 1-m aperture telescope flown on board 
the \textsc{Sunrise} balloon-borne solar observatory
\citep{Solanki:10,Barthol:11,Berkefeld:11,Gandorfer:11}. We present a high-cadence analysis of the
temporal evolution of this magnetic structure. The flux content of the bipole
is $\sim 6 \times 10^{17}$ Mx, which places it close to the limit
between the ERs and the small-scale fields observed at granular scale. In Sect.
2 we describe the observations, in Sect. 3 we report our results on both the
emergence and the decay phases, and in Sect. 4 we discuss our findings in a
general context.

\section{Observations and Data analysis}

\begin{figure}[!t]
\epsscale{1.15}
\plotone{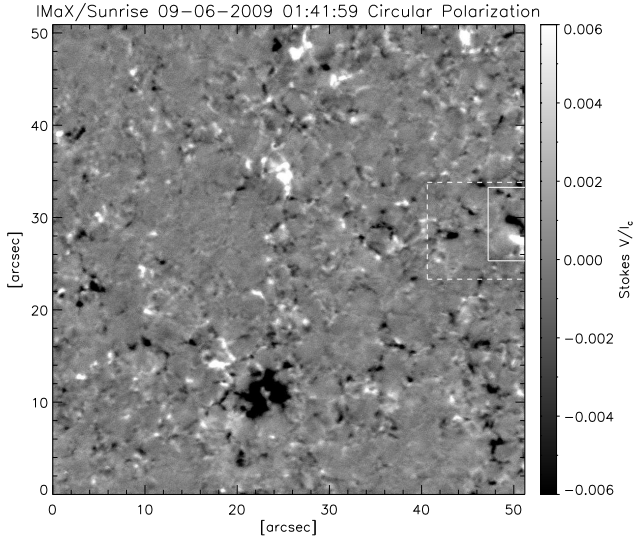}%
\caption{Map of the non-reconstructed mean circular polarization signal $V_s$,
i.e., the LOS magnetic flux, covering the full FoV of IMaX of about $50\arcsec
\times 50\arcsec$, obtained during the observations of the emerging bipole
analyzed in this work. The square (dashed line), with a FoV of $10\farcs5
\times 10\farcs5$, indicates the sub-region whose spectra have been inverted
with the SIR code. The inner rectangle (solid line), with a FoV of $\sim
4\arcsec \times 8\arcsec$ indicates the location of the emerging flux region,
as shown in the sequences displayed in Fig.~\ref{fig3}. \label{fig1}}
\end{figure}

We have analyzed an IMaX data set obtained on 2009 June 9, 01:30:54-02:02:29 UT.
During that time, IMaX took polarization maps at five wavelength positions over
the \ion{Fe}{1} 525.02 nm line (Land\'e factor $g=3$) at $\lambda$ = -8, -4,
+4, +8, and +22.7 pm from the line center, with a spectral resolution of 8.5
pm. Six images were accumulated per wavelength point, and the full Stokes
vector was recorded (V5-6 mode). The temporal cadence of the scans is 33 s, with a pixel size
of 0\farcs055. Figure~\ref{fig1} shows the field of view (FoV) covered by the
observations, about $50\arcsec \times 50\arcsec$ over a quiet region at the
disk center.

All data have been corrected for instrumental effects, by performing dark-current subtraction,
flat-field correction, and cross-talk removal. The blueshift over the
FoV due to the collimated setup of the Fabry-P\'erot etalon of the
magnetograph is corrected in the inferred velocity values. 
Two different types of data are produced:
non-reconstructed data (level 1) and reconstructed data (level 2), obtained by
using phase-diversity information. The reconstruction requires an
apodization that effectively reduces the IMaX FoV down to about $45\arcsec \times 45\arcsec$. 
In this study, we analyze level 1 data, since the emerging bipole was observed just at the 
right border of the full FoV (within the box in Fig.~\ref{fig1}). The spatial resolution 
is 0\farcs3 (before reconstruction) and the noise level is about $1\times 10^{-3}$ in units of 
continuum intensity per wavelength point in each Stokes parameter. Further details about data
reduction are provided by \citet{Valentin:11}.

\begin{figure*}[!t]
\epsscale{1.}
\plotone{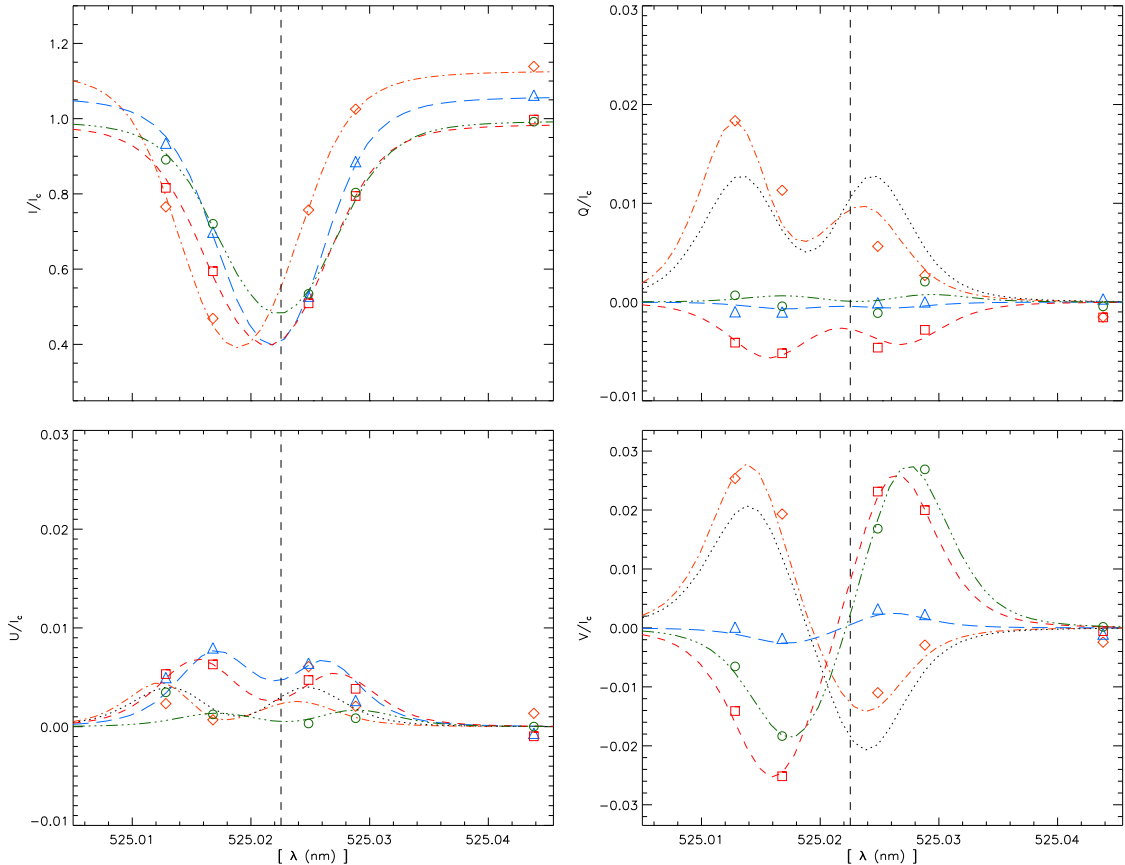}%
\caption{Observed (symbols) $I$, $Q$, $U$, and $V$ profiles, for a sample of
IMaX spectra, with the corresponding fits (lines) obtained with the SIR code.
Blue (triangles/long dashed): horizontal fields, no gradients. Red (squares/dashed):
vertical fields, no gradients. Green (circle/dot-dot-dot dashed): asymmetry in Stokes
\emph{V} with weak $L_s$ signal, with gradients. Orange (diamonds/dot dashed): 
asymmetry in the blue and red lobes of Stokes \emph{U} and \emph{V}, with gradients. 
The dotted line represents the fit without gradients to the plotted orange spectrum (for
Stokes $I$ the dotted line falls on top of the orange one). The
dashed vertical line indicates the nominal line center. \label{fig2}}
\end{figure*}

\begin{figure*}[!t]
\epsscale{1.075}
\plotone{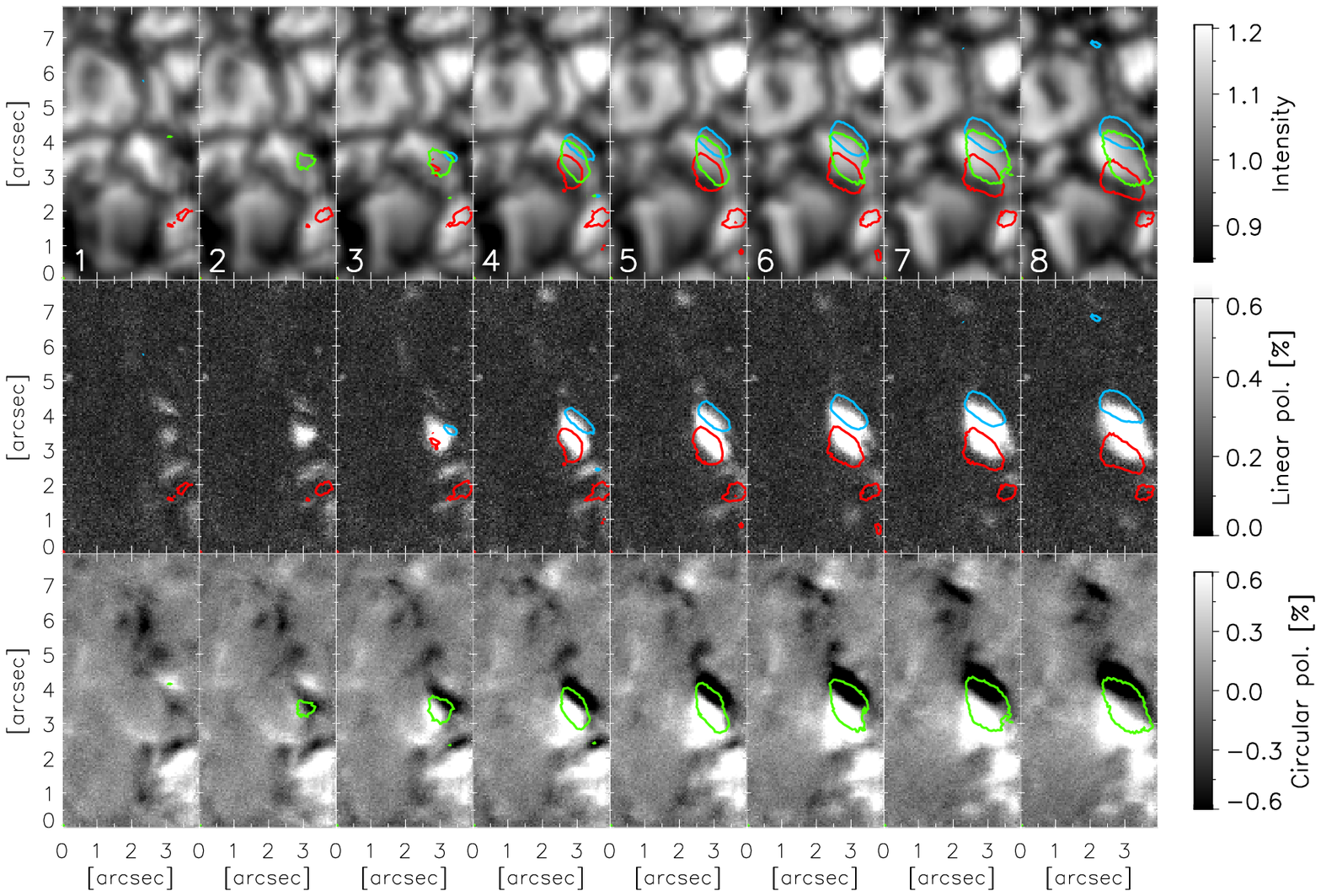}%
\plotone{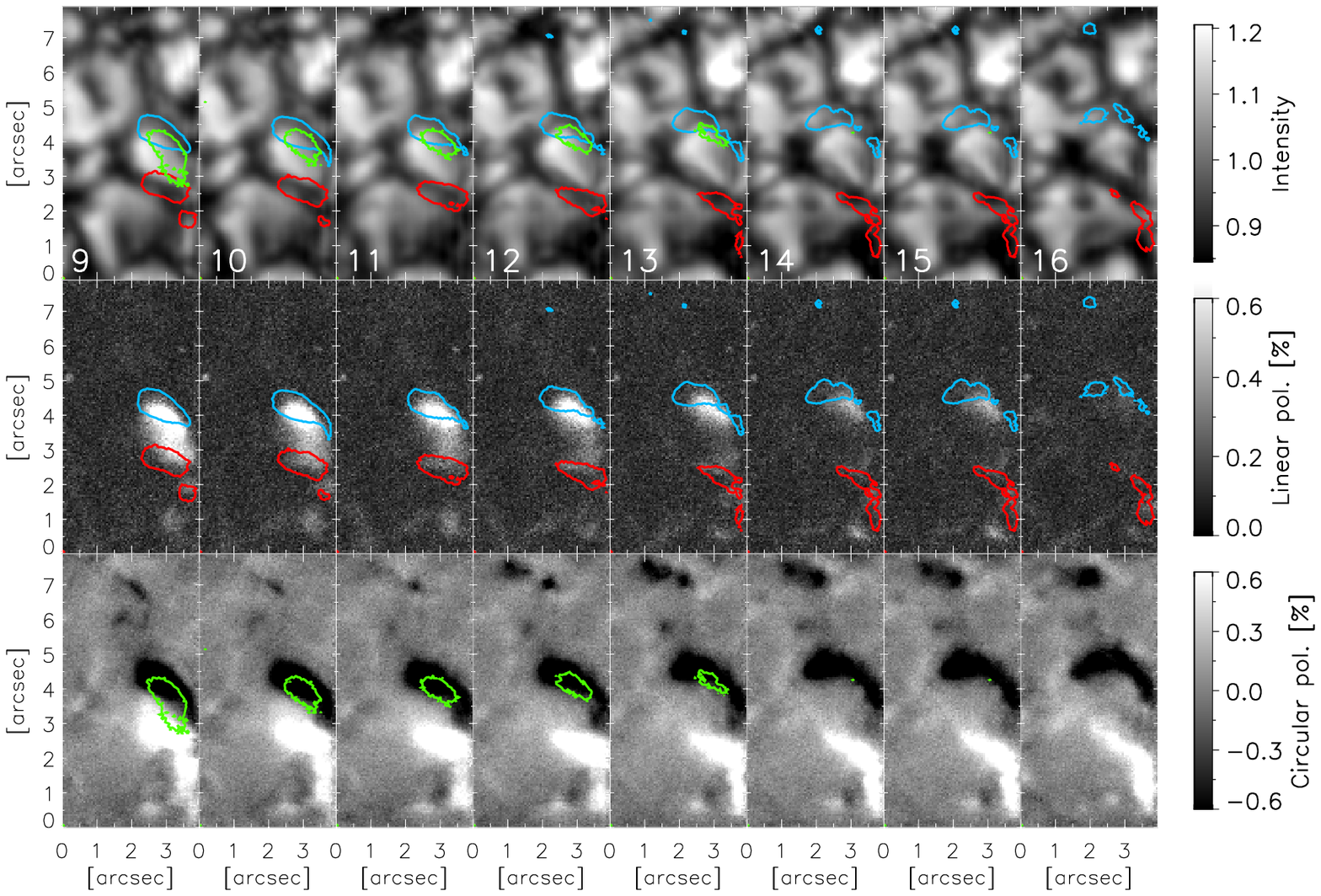}%
\caption{Top rows: Temporal sequence of the continuum maps during the emergence phase 
of the bipole. Time runs from left to right and continues in the lower set of panels.
Red (blue) contours over the  maps represent a circular polarization signal of +0.8 (-0.8) \% of the $I_c$,
and green contours represent a linear polarization signal of 0.5 \% of the $I_c$. Middle rows:
the same for the mean linear polarization signal $L_s$. Bottom rows: the same for the total 
circular polarization $V_s$.
The elapsed time between consecutive frames is 33 s. From frame 11 the decay phase begins.
An animation of this figure is available in the online journal. \label{fig3}}
\end{figure*} 

We obtained maps of the mean circular polarization averaged over the line,
$V_s$, and of the mean linear polarization signal $L_s$, given respectively by

$$V_s=\frac{1}{4\left\langle I_c\right\rangle} \sum_{i=1}^{4} \epsilon_{i} \left|V_i\right|$$

\noindent
and 

$$L_s=\frac{1}{4\left\langle I_c\right\rangle} \sum_{i=1}^{4}\sqrt{{Q_i}^2+{U_i}^2}\,,$$ 

\noindent
where $\left\langle I_c\right\rangle$
is the continuum intensity averaged over the IMaX FoV, $\epsilon =[1,1,-1,-1]$, 
and $i$ runs over the first four wavelength positions. 

We carried out inversions of the observed \ion{Fe}{1} 525.02 nm Stokes
vector spectra using the SIR code \citep{RuizIniesta:92}. This code numerically
solves the radiative transfer equation along the line of sight (LOS) for
Zeeman-polarized radiation under the assumption of local thermodynamic
equilibrium.

The inversion yields the temperature stratification in the range $-4.0 < \log
\, \tau < 0$, where $\tau$ is the optical depth of the continuum at 500 nm. 
We use the Harvard Smithsonian Reference Atmosphere \citep[HSRA;][]{Gingerich:71}
as initial model. The temperature is modified with two nodes.
SIR also provides the LOS velocity $v_{LOS}$, the
micro-turbulent velocity $v_{micro}$, the magnetic field strength $B$, the
inclination and azimuth angles $\gamma$ and $\phi$ in the LOS reference frame.
As these IMaX observations were taken at disk center, the returned magnetic parameters 
do not need to be converted to local solar coordinates. The magnetic filling factor 
has been assumed unity for these inversions, not taking into account stray light contamination. 
The synthetic profiles are convolved with the spectral Point Spread Function at the focal plane of IMaX.

In a first series of inversions, all the physical parameters are assumed to be constant with height,
except for the temperature that has two nodes, as already mentioned above. In another series of inversions,
$v_{LOS}$ and $B$ are also allowed to have a linear dependence with the
optical depth $\tau$. Such a gradient in both parameters has been found to
be necessary in order to fit some profiles that show a considerable asymmetry
between the blue and red lobes of Stokes $V$, or between the lobes of $Q$
and $U$ when they are significant. Elsewhere the fits obtained without gradient
in velocity and magnetic field strength is found to be as good as the one with
gradients. The total number of free parameters is 7 for SIR inversions (9 if 
gradients in $v_{LOS}$ and $B$ are included) and the number of the data points
in the V5-6 mode is 20.

The absolute velocity is calibrated using the results coming from the Gaussian
fits obtained for the same data set \citep[see, e.g.,][]{Roth:10}, compared
with the results of the SIR inversion without gradients. This calibration uses
a granular mean blueshift of 200 m s$^{-1}$.

In Fig.~\ref{fig2} we display a subset of the typical profiles acquired by
IMaX, and the corresponding fits obtained using the SIR code. The spectra
plotted in blue, showing predominantly horizontal fields, and red, with more
vertical fields, show profiles well fitted without any gradient.  There are
also clear examples of spectra showing noticeable asymmetries in Stokes $V$ in
absence of a significant $L_s$ signal (plotted green), as well as in Stokes $Q$
and $V$ with clear horizontal fields (plotted orange), which both require a
gradient in $v_{LOS}$ and $B$. These profiles show either downflows or upflows,
respectively, as can be deduced also from the shifts with respect to the line
center in Stokes $I$. The residuals of the fit, defined as $P_{obs}-P_{fit}$,
with $P$ being any of the polarization profiles, show an rms variation over the
inverted FoV of $1-1.6 \times 10^{-3}$, in agreement with the estimated noise
per wavelength sample of the data.

\section{Results}

\subsection{Flux emergence phase}

Figure~\ref{fig3} shows a temporal sequence of the continuum intensity
$I_c$, and of the polarization maps $L_s$ and $V_s$. This sequence tracks the
emergence of the bipole since its first detection in the
photosphere. Green contours enclose the region with a significant linear
polarization signal, i.e. $L_s > 0.5\%$, whereas red and blue contours mark
areas with positive and negative circular polarization, respectively, 
with $V_s > 0.8\%$. We also provide a movie of the evolution of the bipole in the online journal.
The temporal cadence of each frame is 33 s.

At 01:35:54 UT the bipole appears in frame 2 at coordinates [3\arcsec,3\farcs5], 
inside a pre-existing granule visible at the same location in frame 1.
It is later recognizable as a patch with increasing linear polarization signals
in the following $L_s$ maps that always coincides with the top of the granule
during the whole emergence process. There is no evidence of any influence of
the emerging magnetic flux on the granulation pattern.

Note that the bipole emerges in a region where two subarcsecond
bipoles were already present as seen in the first $V_s$
frame of Fig.~\ref{fig3} (top panel). One of these bipoles
is located at [3\arcsec,4\arcsec], nearly exactly over the region of emergence of the negative 
footpoint of our bipole, but with its polarities flipped (i.e., with the positive 
polarity toward the top part of the figure) with respect to those of the emerging bipole. 
The other is located at the bottom right  of the studied area, at [3\farcs5,2\arcsec],
and possesses the same orientation as our bipole. 

The corresponding $V_s$ footpoints of the
large emerging bipole are first seen in the frames 3 -- 4, as two small patches of
opposite polarities at symmetrically opposite edges of the enhanced $L_s$ patches. 
The evolution of the emerging loop can be easily
followed in the subsequent $V_s$ frames, where the opposite polarities of the
bipole, i.e., the footpoints of the loop, are seen to separate from each other in
opposite directions, with quite a strong $L_s$ signal in between them 
being maintained until frame 9. This implies horizontal field in mid-photospheric layers, which is
consistent with continuing flux emergence during this whole period of time.

It is clear that the emerging bipole studied in this work interacts with the two pre-existing ones.
The first bipole at [3\arcsec,4\arcsec] is completely overtaken by the negative polarity footpoint
of our larger emerging bipole. The positive polarity footpoint of the new bipole, however,
first enters an intergranular region but crosses over it to interact with the negative polarity
footpoint of the pre-existing bipole (frame 5 in Fig.~\ref{fig3} at [3\farcs5,2\farcs5]).
The negative pre-existing footpoint is completely washed away, while the emerging positive 
footpoint appears to aggregate the positive polarity flux of the pre-existing bipole.

\begin{figure*}[!t]
\epsscale{1.1}
\plottwo{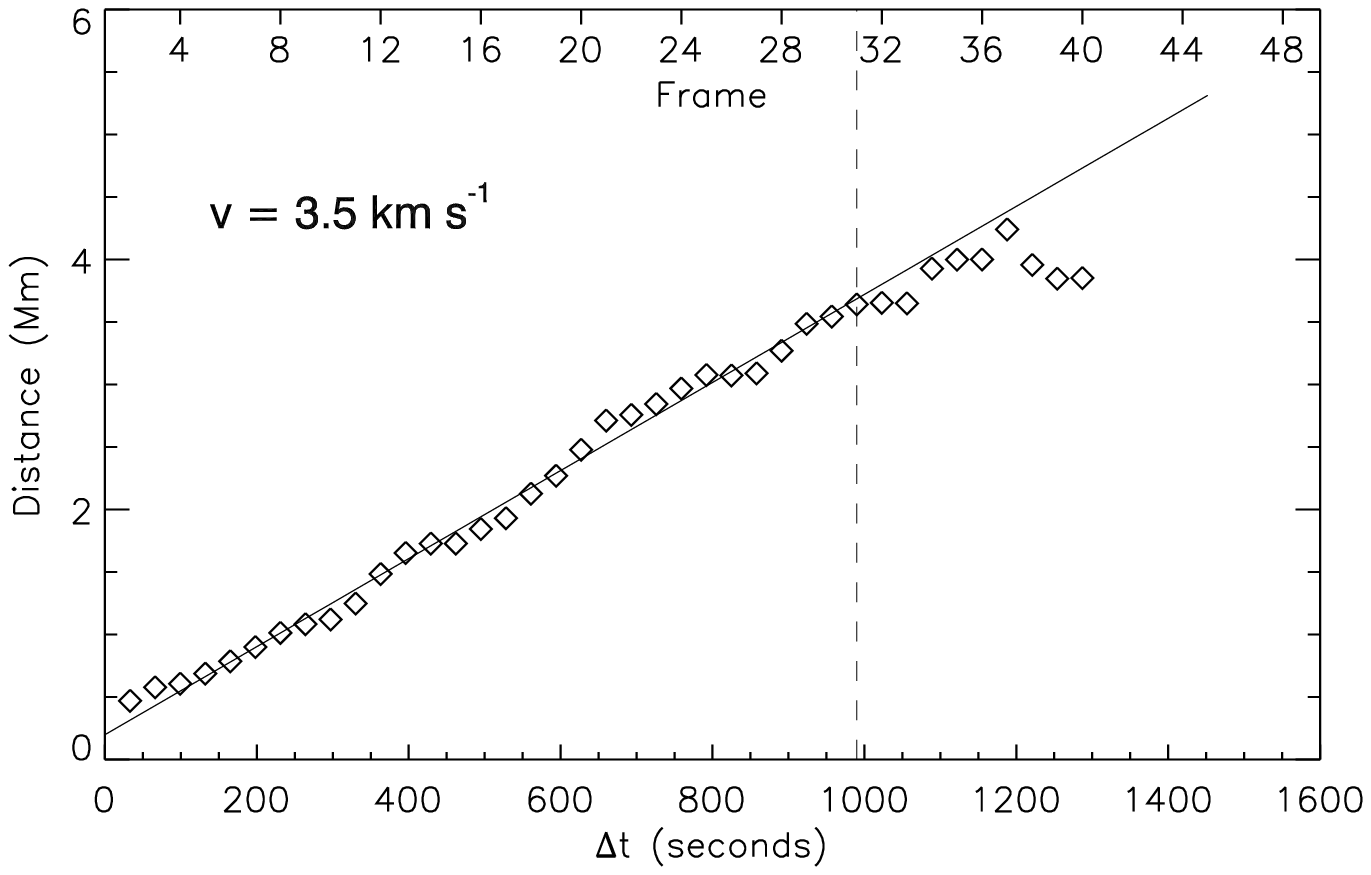}{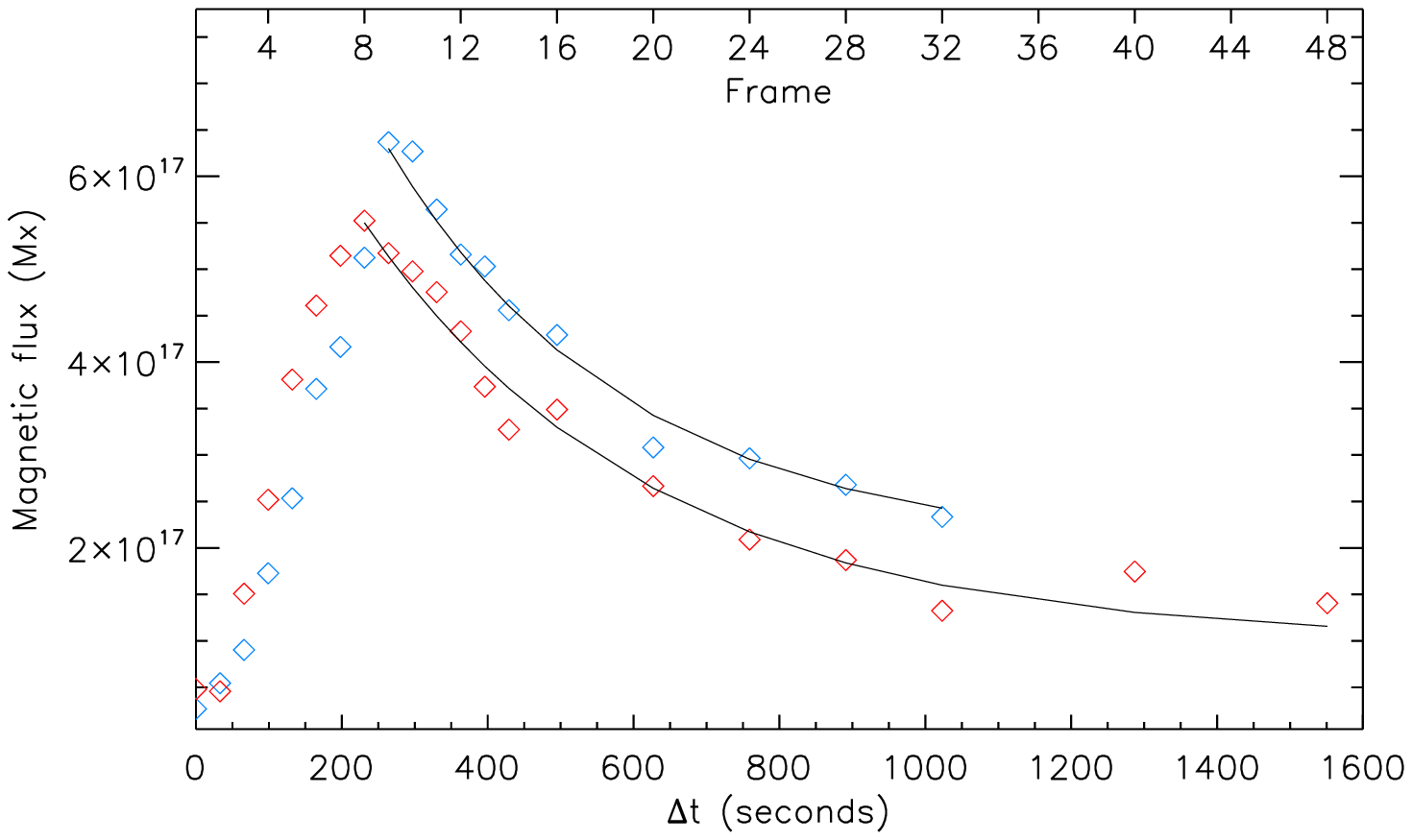}%
\caption{Left panel: Distance between the centroids of the opposite polarity footpoint 
of the emerging bipole as a function of time. The dotted vertical line indicates the time
at which separation deviates from a linear increase. Right panel: Evolution of the flux 
content in the positive (negative) polarity of the emerging bipole. Red (blue) symbols indicate data
values, solid lines represent offset exponential fits to data during the decay phase. \label{fig4}}
\end{figure*}

We plot the distance between the footpoints as a function of time in
Fig.~\ref{fig4} (\emph{left panel}), calculated as the distance between the centroids
of the areas where $\left|V_s\right|$ is greater than 0.5\%.
The plot displayed in Fig.~\ref{fig4} (\emph{left panel}) shows that the distance increases linearly
during the first 1000 s of the loop evolution, with a mean
separation velocity of 3.5 \kms. Maximum footpoint separation is 4.5 Mm.
This result has to be compared with the analysis of emerging bipoles 
carried out by \citet{Marian:09}. In some cases, they found an initial velocity separation 
of the footpoints of 6 \kms, with a change after 500 s to $\approx 2$ \kms.
Note that our bipole is comparable to the largest bipoles studied in that work. 

\begin{figure*}[!t]
\epsscale{1.075}
\plotone{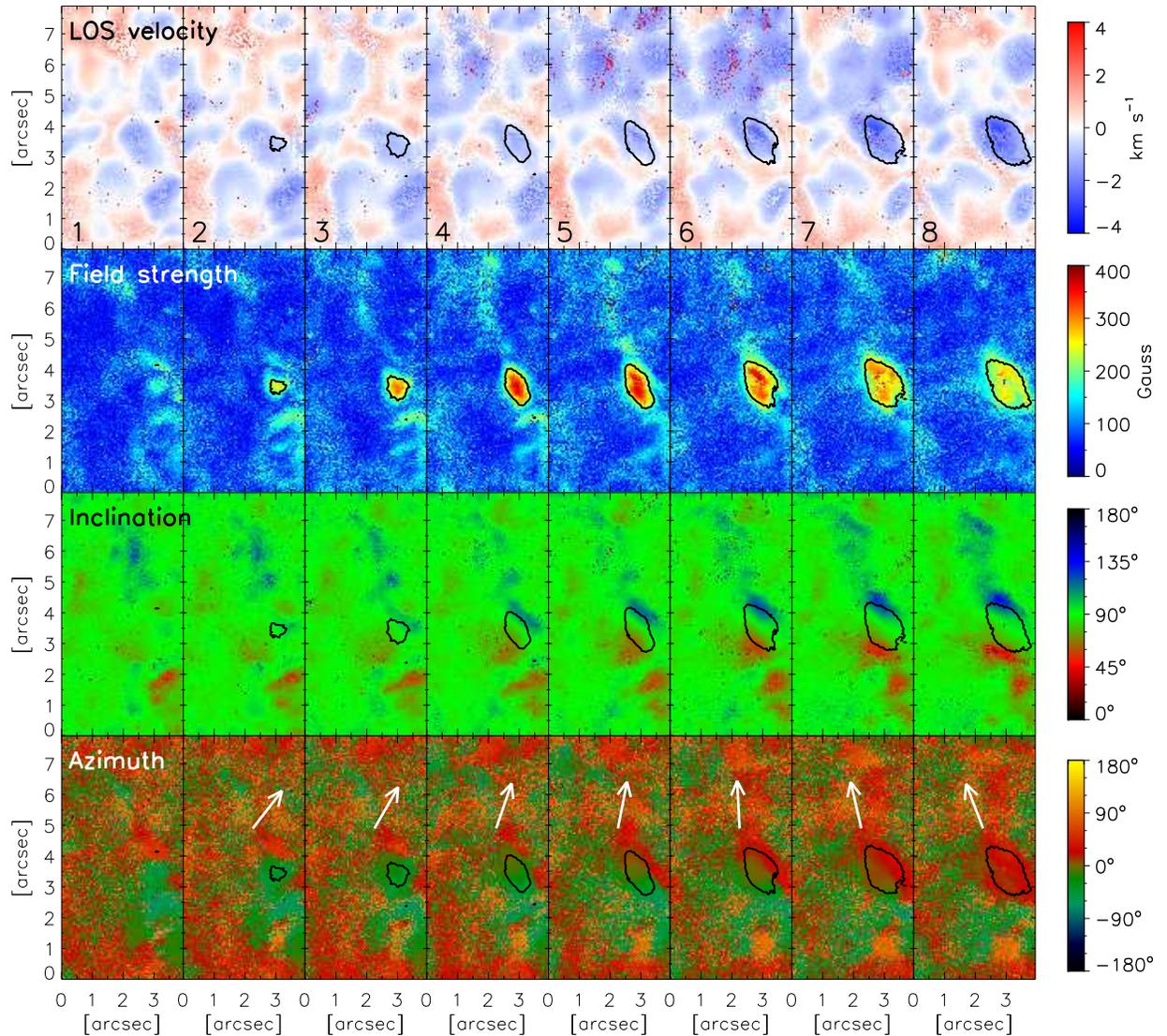}%
\caption{LOS velocity and magnetic field strength at $\log \, \tau = -2$ (first
and second rows), inclination and azimuth angles (third and fourth rows), for
frames 1 -- 8 (compare with Fig.~\ref{fig3}). Contours represent a linear polarization signal of 0.5 \% of the
$I_c$, as in Fig.~\ref{fig3}. Arrows indicate the mean value of azimuth angle $\phi$ within the $L_s$ countour for each frame. 
The direction to the top of the plot is assumed as zero reference for $\phi$. \label{fig5}}
\end{figure*}

\begin{figure*}[!t]
\epsscale{1.075}
\plotone{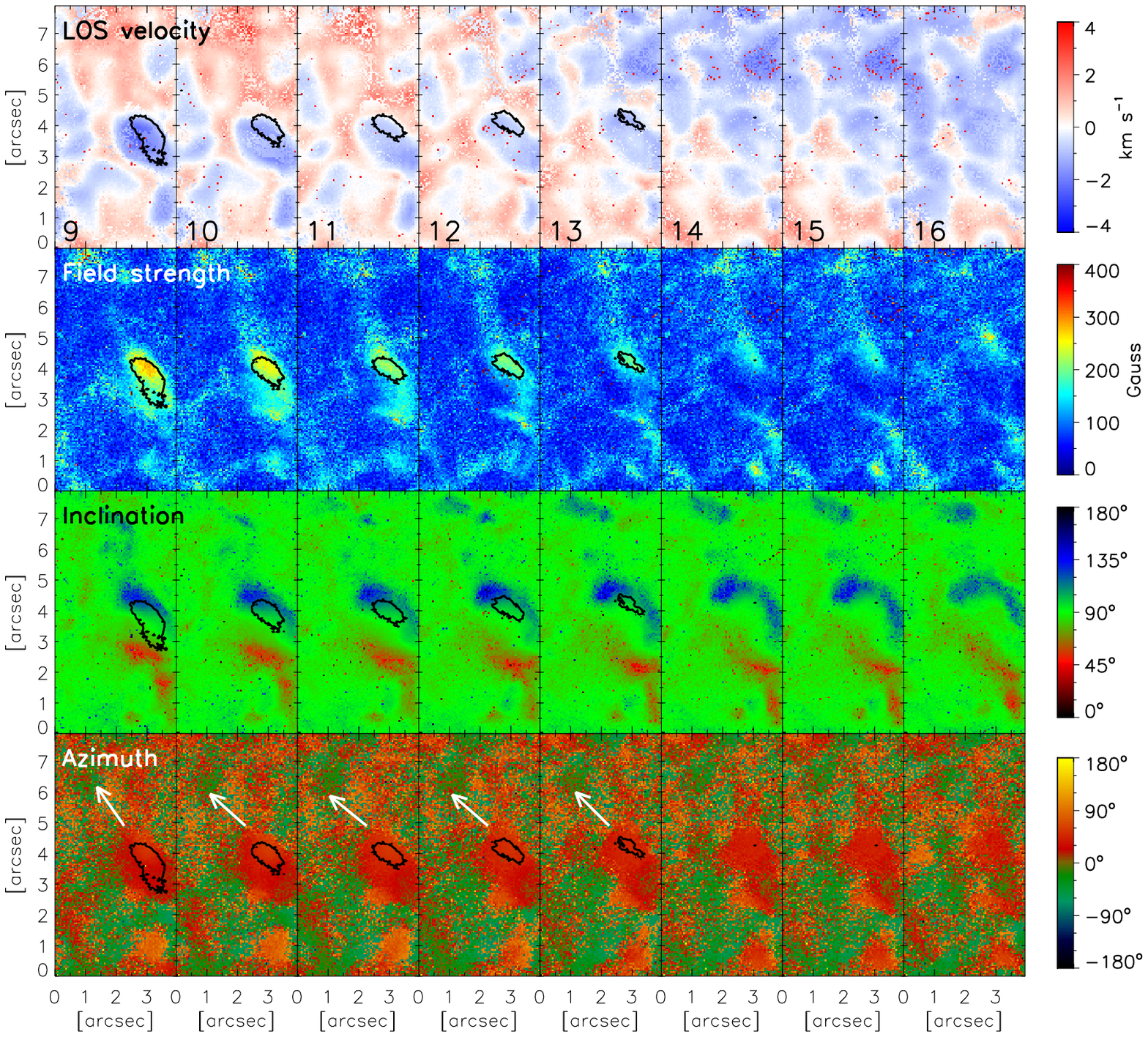}%
\caption{Same as in Fig.~\ref{fig5}, for frames 9 -- 16. \label{fig6}}
\end{figure*}

The corresponding temporal sequences of the physical parameters retrieved by
SIR during the emergence are shown in Fig.~\ref{fig5}, for frames 1 -- 8 of Fig.~\ref{fig3}, and
in Fig.~\ref{fig6}, for frames 9 -- 16. We display the values of LOS velocity and
of the magnetic field strength averaged between $\log \, \tau = -1$ and $\log \, \tau = -2$, where the response
functions are more sensitive (note that no filling factor is included in the
inversions). Blueshifts, which indicate upward motions, are present in the
region since the beginning of the emergence until frame 10. They are cospatial
with the region with strong $L_s$ signals, indicative of nearby horizontal fields,
and correspond to the granule in the continuum map. The emerging flux region is
surrounded by downflows along the intergranular lanes which outline the granule. 
The larger scale variable velocity pattern near the top of the velocity frames 5 -- 7
reflects the \emph{p}-mode oscillations, as discussed by, e.g., \citet{Roth:10}.

The magnetic field strength has a peak of about 400 G around frames 5 -- 6,
then it begins to fade (note that this occurs before the footpoints enter the
intergranular lanes, which happens in frame 7). As our filling factor is unity,
this value should be taken as a flux-equivalent field strength. The maximum
values of $B$ are found in the region where the magnetic field is more
horizontal. These strengths are of the order of the typical equipartition field
strength $B_{e}$ for granules, given by 

$$\frac{{B_{e}}^{2}}{2\,\mu_{0}}=\frac{1}{2}\rho \,v\,,$$

\noindent
where $\mu_{0}$ is the magnetic permeability of free space, with 
$v \approx 2$ \kms{} and $\rho = 3 \times 10^{-4} \;\textrm{kg m}^{-3}$
\citep{Ishikawa:08}. 

The footpoints do not reach a vertical orientation, but they remain rather inclined 
at an angle of about $45^{\circ}$ with respect to the vertical. Furthermore, they surround the
granule having a crescent-shaped aspect, different from the usual circular shape found in previous observations.

In the emergence zone, the azimuth angle is quite homogeneous at every instant during the rise,
with the scatter ranging from $\pm 5^{\circ}$ to $\pm 15^{\circ}$ around the average value. However,
the mean value, indicated by arrows in Figs.~\ref{fig5} and~\ref{fig6},
changes with time, rotating \emph{counterclockwise} $10^{\circ}$ per frame until the end of
the emergence\footnote{Note here that we do not solve the azimuth ambiguity.
The arrow is selected simply by choosing the azimuth direction that points 
from the positive to the negative footpoint of the bipole}.
The observed rotation rate is thus $0^{\circ}$.3 s$^{-1}$. In the $L_s$ and $V_s$ maps in Fig.~\ref{fig3}, 
we can see that the bipole axis also rotates in the same \emph{counterclockwise} direction. 
This similar rotation for both the field azimuth angle and the bipole footpoints is compatible with 
the absence of writhe in the emerging flux tube. The observed evolution of the azimuth, which steadily varies 
by $\Delta \phi \approx 90^{\circ}$ during the emergence, may be attributed to convective motions, 
but we cannot rule out the presence of some twist in the structure, unfolding during its ascent.

The total magnetic flux content is given by

$$\Phi = \sum_{i} B_{i}  \cos \gamma_{i} \,S \;,$$ 

\noindent
where index \emph{i} runs over all the pixels of the bipole with the same
polarity and \emph{S} is the area of the Sun covered by a pixel. 
We note here that our assumption of a filling factor equal unity, while probably affects
the values found for the magnetic field strength, has no
impact on the estimate of the magnetic flux density and on the total flux
content. In both polarities, the flux grows linearly until a
maximum of about $6 \times 10^{17}$ Mx at frame 8 is reached, 4 minutes after
the first detection of the bipole as seen in Fig.~\ref{fig4} (\emph{right panel}).
In this figure, the fluxes were estimated by adding all the contributions from
manually selected pixels in the footpoint area of interest. Missing points correspond
to non-inverted frames. The flux growth rate is thus $2.6 \times 10^{15}$ Mx s$^{-1}$.

Note that the slightly smaller positive flux (red symbols) may be due to
a different inclination between the footpoints and, as footpoints move apart, to
the fact that the corresponding footpoint is located closer to the right
boundary of the observed area. At the end of the emergence process (defined as
the maximum in the flux curve), the distance between the footpoints of the
bipole is about 1.5 Mm, with roughly the same size as a typical granule. All
these results are consistent with the emergence of an $\Omega$-shaped magnetic
bipole at granular scale.

The maximum distance \emph{D} attained by the footpoints, 4.5 Mm, is roughly in agreement
with the linear $\log D / \log \Phi$ relation found by \citet{Hag:01} for ERs, which
gives $D \simeq 3.5 \times \Phi^{0.18} = 5.5$ Mm.

\begin{figure*}[!t]
\epsscale{1.15}
\plotone{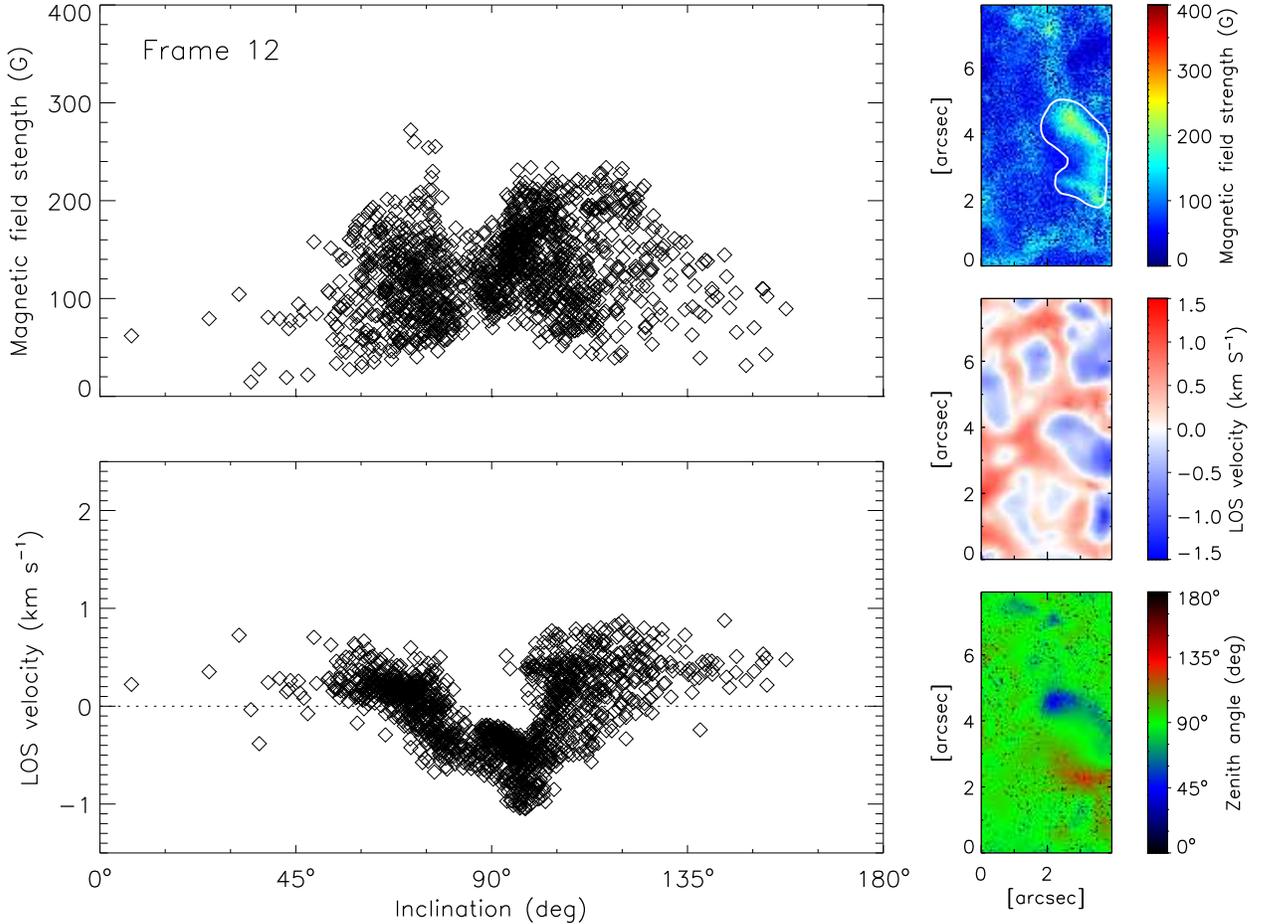}%
\caption{Scatter plots of $B$ and $v_{LOS}$, as a function of the inclination
$\gamma$, for a representative frame (No. 12 in Fig.~\ref{fig3}). We show the results for the SIR inversion
without gradients, and the corresponding maps of the physical parameters. Only the pixels within the white contour 
in the $B$ map with a signal than three times the noise have been indicated in the plot. Keep in mind that inversion have a tendency to retrieve horizontal fields for pixels at noise level. 
\label{fig7}}
\end{figure*}

Active region flux emergence displays upflows associated with transverse fields
at equipartition values and downflows at the footpoints with much
larger field strengths \citep[see, e.g.,][]{Lites:98,Solanki:03}. In order to compare this
behaviour with our bipolar emergence, we have produced a scatter plot of the
magnetic field strength and LOS velocity vs. zenith angle $\gamma$, as shown in
Fig.~\ref{fig7} for the SIR inversion without gradients in B and $v_{LOS}$ (frame 12).
In that frame, the footpoints of the loop, which correspond to the
regions with more vertical field, have also stronger field strengths of
about $200 - 250$ G with downflows slightly larger than 1 \kms, decreasing with time. As
already stated above, it is conceivable that these fields are indeed at
equipartition values. In the horizontal field regions, an upflow of 1.0 \kms{} is seen.
This upward motion also decreases with time. Peak upflows for the transverse
fields are seen in frames 7 -- 9 where they reach 2.5 \kms. 

No evidence of asymmetry between the footpoints of the bipole is found in the data in both the
field strength and the LOS velocity. However, the linear polarization signal, $L_s$ is concentrated
mainly at the negative footpoint in frames 10 -- 13.

\subsection{Flux decay phase}

The observations used in this work allow a detailed analysis
of the decay phase of the bipole, the less known stage
of the evolution of small bipoles and ephemeral regions.
The footpoints of the emerged bipole remain visible until the
end of the emergence process and well beyond into the decay phase. This
phase can be defined to start at the time when the total measured
flux begins to decrease, i.e., from frame 9 onwards.
Note that the flux curves in Fig.~\ref{fig4} (\emph{right panel}) never reach a flat portion 
where the flux of the footpoints would stay constant. Instead, as soon as the 
maximum is reached, a mechanism somehow begins to erode the newly emerged 
flux. 

\begin{figure*}[!p]
\epsscale{1.05}
\plotone{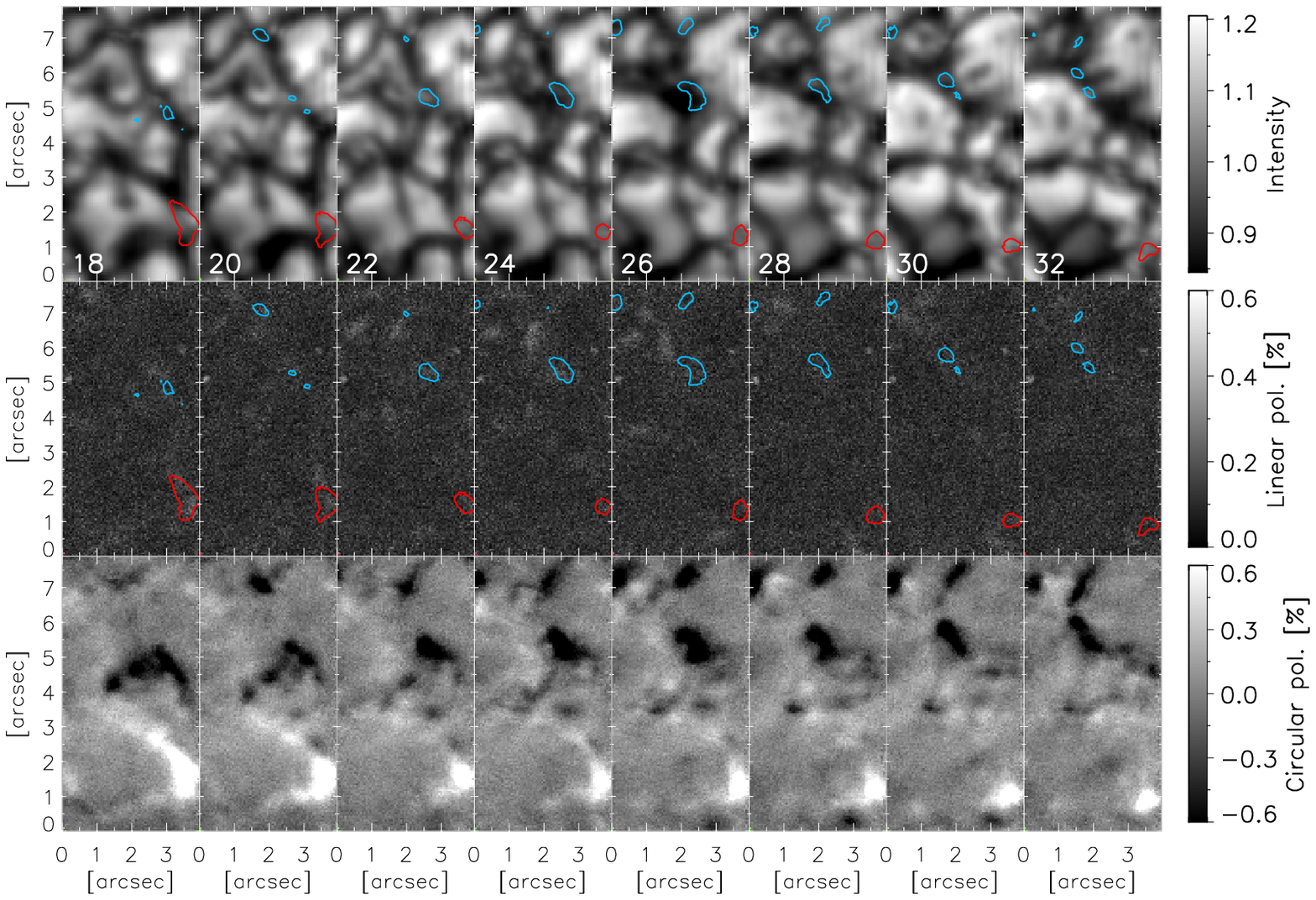}%
\plotone{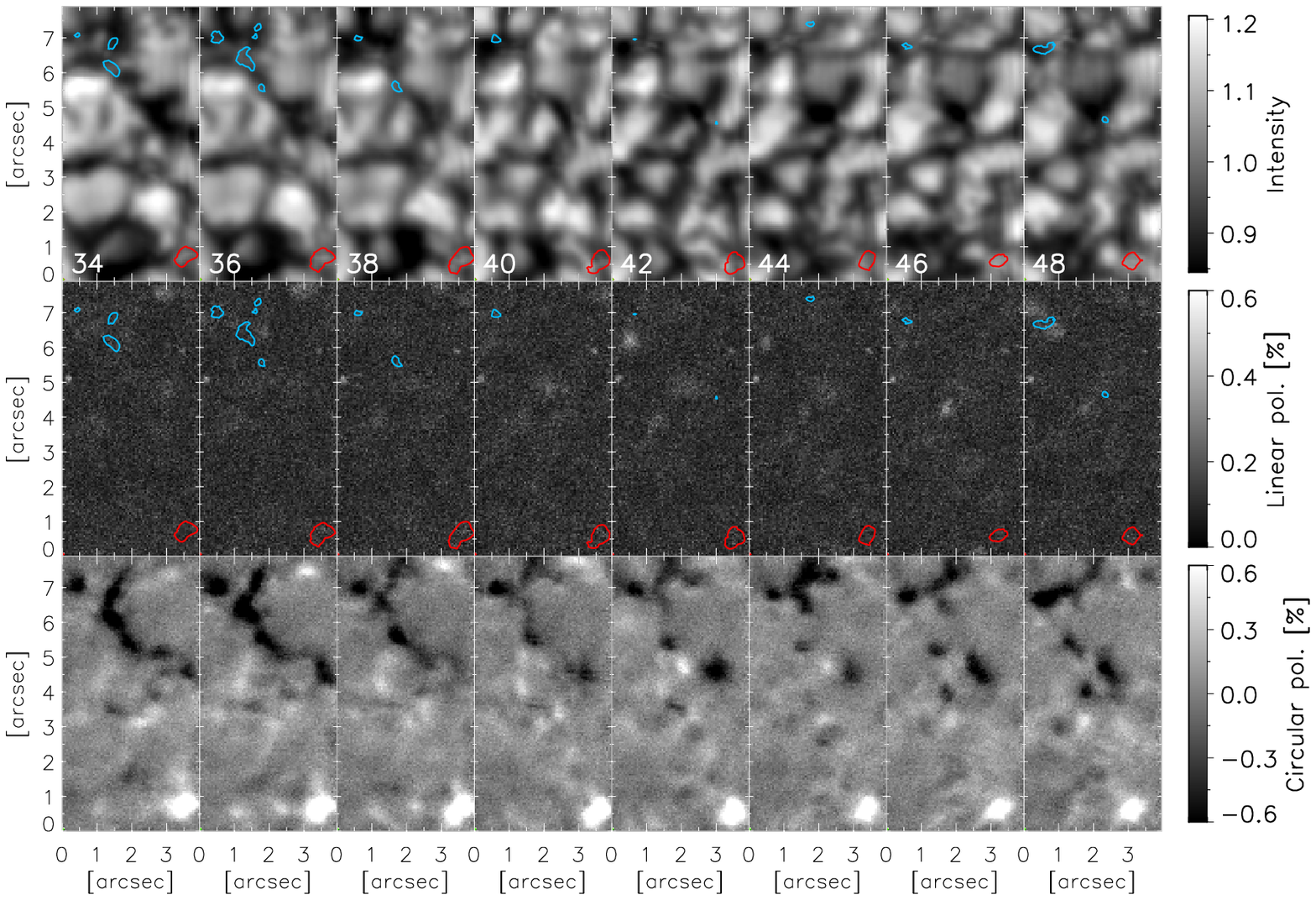}%
\caption{Same as in Fig.~\ref{fig3}, but during the decay phase of the bipole.
Frame numbering continues from the last frame of Fig.~\ref{fig3}. In this
figure, the elapsed time between consecutive frames is 66 s. \label{fig8}}
\end{figure*}

Figure~\ref{fig8}, which is the continuation of Fig.~\ref{fig3}, shows the evolution 
of the bipole after the emergence process: since the evolution during this phase is 
less rapid, only every other frame is shown (but see also the on-line material). 
Footpoints move apart until they start breaking into smaller flux patches 
and interact with the surrounding fields. Their remnants remain distinguishable 
at least for 15 minutes after the first detection of the magnetic structure. 
In frames 16 -- 18, the bulk of the negative polarity starts being fragmented, losing its compactness. 
During this process the various patches of this polarity move along intergranular lanes.  

In frames 32 -- 34 the negative polarity of the bipole totally
merges into a pre-existing flux element of the same polarity, and is no
longer distinguishable as an individual flux concentration. This explains why the
flux curve for the negative polarity footpoint in the right panel of Fig.~\ref{fig4} stops there. 

The positive polarity footpoint at the beginning of the decay phase has merged with the pre-existing 
positive flux of the bipole at the bottom. Note that, in this case, no net flux
is lost as the positive polarity footpoint of the pre-existing bipole
is aggregated to the positive polarity of the newly emerged one. This could
explain why this cancellation episode is not apparent in the red flux
curve of Fig.~\ref{fig4} (\emph{right panel}). After this cancellation, the newly
created positive polarity patch is seen to move near the edge of our field
of view with little evident interaction with other fluxes (see Fig.~\ref{fig8}).
Interestingly, during this time the positive flux patch moves largely over granulation
dominated regions instead of intergranular ones.

In order to get an indication of the physical processes occurring during the
decay phase, we have fitted the flux evolution curves during the decay period
using the MPFIT routine \citep{MPFIT}, which performs fits to data with user-defined functions. 
We have used as a guess function an exponential of the type 

$$f\left(t\right) = a + b \cdot e^{\beta\,t} \,\,\,.$$ 

We obtained $\beta_{\pm}=-2.5 \times 10^{-3}\;\textrm{s}^{-1}$ and $-3.0 \times
10^{-3}\;\textrm{s}^{-1}$ for the positive and negative flux, respectively.
This corresponds to a ``turbulent'' diffusion costant $\eta = L^{2}/\tau$ of
the order of $8 \times 10^{2} \;\textrm{km}^{2}~\textrm{s}^{-1}$, where $\tau
= \beta^{-1}$ is the \emph{e}-folding time and $L \approx 0\farcs8$ is the size of the footpoint at
the time of the flux peak in Fig.~\ref{fig4} (\emph{right panel}). It is
interesting to note that this diffusion constant compares rather well with
those inferred by flux transport models of the large scale field, typically 
set at $6 \times 10^{2} \;\textrm{km}^{2}~\textrm{s}^{-1}$ \citep[see
the instructive review by][]{Sheeley:05}. 

This order of magnitude agreement between the diffusion constants of the large scale 
field and of our granular scale bipole is rather illuminating. As the flux loss occurs
\emph{in situ}, at a rate similar to that encountered in flux transport models, 
this could indicate an analogous physical process acting in both flux loss scenarios.

Even though the flux decay curve is clearly non-linear, it is
interesting to compute a mean flux decay rate which turns out 
to be $2.86 \times 10^{14}$ Mx s$^{-1}$. This rate
is one order of magnitude smaller than the flux growth rate. A similar
asymmetry between the growth and decay rates is known to exist for active 
regions \citep{Valentin:02,vanDriel:02}.

\section{Discussion and Conclusions}

IMaX/Sunrise observations of the solar photosphere taken at disk center have
revealed a number of small-scale episodes of magnetic flux emergence
\citep[see, e.g.,][and references therein]{Danilovic:10,Solanki:10}.  We have
carried out an analysis of the emergence and disappearance of a small magnetic
bipole. We have analyzed the polarization maps and then we have inverted the
Stokes profiles with the SIR code, to obtain information on the physical
parameters of this magnetic structure. 

The detection of weak field strengths (400 G) at emergence with horizontal
inclination, associated with blueshifted Stokes line profiles between opposite
polarity Stokes V profiles, is a well-known signature of the emergence of a
concentrated flux loop, as firstly pointed out by \citet{Lites:96} at active
region scales.  At smaller (granular) scales, our analysis shows results
resembling those found by \citet{Gomory:10}, but with a higher cadence and
better spatial resolution. 

The magnetic flux content places this small bipole at a halfway point between
the ephemeral regions studied by \citet{Hag:01} and the all-pervasive loops of
the quiet Sun. The bipole indeed appears to be coincident with a granule in the
continuum map, so it would apparently represent a typical case of flux
emergence at granular scale. However, the present case is the
largest structure observed to emerge in the IMaX/Sunrise data analyzed so far.

Once emerged, the magnetic flux spreads over an area larger than the source
granule. The footpoints separate up to 4.5 Mm from each other and their remnants remain
visible until 15 -- 20 minutes after the first detection of the magnetic structure.
Nevertheless, the granulation pattern seems not to be affected by the flux
brought to the surface by the emerging bipole. This agrees with the simulations 
of \citet{Cheung:07}, showing that flux tubes carrying less than $\sim 10^{18}$ Mx
of longitudinal flux do not produce visible disturbances to the granulation pattern.

During the emergence, the footpoints of the bipole do not have the more typical
circular shape seen in Hinode observations \citep[e.g.,][]{Marian:09}.  They
rather surround the granule and have a crescent-shaped appearance.  Moreover,
footpoints do not reach a vertical orientation: they form an angle of about
$45^{\circ}$.  This could suggest that only a part of the $\Omega$-loop is
observed above the continuum formation layer, as also found by
\citet{Ishikawa:10}, or that the field lines follow the borders of the granule,
which could not be vertically steep.  Another possibility would be that the
field in the footpoint of this emerging loop is not resolved at the spatial
resolution ot IMaX level 1 data. Former measurements presented by
\citet{Solanki:96} showed that quiet-Sun feature with an amount of flux over
$10^{17}$ Mx tend to have kG field strength, while the inversions retrieve a
field strength of $\approx 250$ G. An underestimate of the field strength
implies that the inclination angle $\gamma$ is overestimated
\citep{Stenflo:85a,Stenflo:85b}, so that the magnetic field in the footpoints
would be more vertical than the inversions indicate. Anyway, such a difference
of more than a factor of 3 in the field strength is rather difficult to
conceive for our measurements based on the high Zeeman sensitive \ion{Fe}{1}
525.02 nm line ($g=3$) displaying Stokes $Q$ and $U$ signals above the
noise level (see, e.g., frame 12 in Fig.~\ref{fig3}). These signals help inferring 
the field strength in the inversions more reliably than when only Stokes $V$ is used.

The azimuth angle changes during the emergence by about $90^{\circ}$ (see 
Figs.~\ref{fig5} and~\ref{fig6}), in the same direction as the line joining the 
footpoints. Convective motions acting on the emerging loop or the unfolding
of any twist present in the original structure could generate this behaviour.

The flux decay follows an exponential law. The observed cancellation and
merging with a pre-existing bipole does not have an effect on the flux history
of the bipole and is not the dominant decay process in this example, although
it might change the connectivity of the flux patches.  Flux removal dominates
from the moment in which the maximum flux is observed. Clearly, flux is
destroyed \emph{in situ}, i.e., without visible interactions with opposite
polarity patches, in some efficient way that dominates the flux budget as soon
as no more flux is brought to the surface. This elusive process probably acts
also during the flux emergence phase, but is not evident at this stage simply
because emergence is one order of magnitude stronger than the flux decay rate.
The diffusion constant estimated for this process is only a little bit higher
than typical estimates used for large scale flux transport models ($6 \times
10^{2} \;\textrm{km}^{2}~\textrm{s}^{-1}$). 
\citet{Cameron:11} have also found similar values in numerical 
simulations of small-scale mixed polarity field in the near-surface layers of 
the Sun. They show that magnetic elements are advected by the horizontal granular 
motions against each other and that the flux is removed at a rate corresponding
to an effective turbulent diffusivity of about $3 \times 10^{2} 
\;\textrm{km}^{2}~\textrm{s}^{-1}$, which is only slightly smaller than our estimate.

The existence of flux-removing diffusive processes that are similarly effective at active 
region and granular scales deserves to be further studied from a theoretical and observational
point of view. Our value of the diffusion constant can be used to estimate the time in 
which an active region would decay. As \citet{Meyer:74} inferred for the turbulent diffusion 
of a monolithic magnetic flux concentration, the decay time is given by

$$t = \frac{\Phi}{4\,\pi\,\eta\,B_{e}}\,,$$

\noindent 
where $\Phi$ is the original flux content, $\eta$ the turbulent diffusivity, and $B_{e}$ the
equipartition field strength. Using this formula for a typical active region of 10$^{22}$ Mx
with our value of the turbulent diffusivity ($8 \times 10^{2} \;\textrm{km}^{2}~\textrm{s}^{-1}$),
we estimate a decay time of $\approx 3$ days. As active regions live longer, it seems clear that 
the diffusivity found in this paper is more effective on smaller flux structures, or possibly 
on flux structures embedded in the vigorous granular convection found in the quiet Sun, 
rather than in the smaller and more slowly evolving abnormal granulation in active regions. 

Finally, it is interesting to compare the emergence rate with the distribution
of emerging fluxes proposed by \citet{Thornton:11}. The number of events per
unit area per unit time within a given flux range $\Phi$ is 

$$N_{\rm ev}\left(\Phi\right) =
\frac{n_0}{2-\alpha}\left(\frac{\Phi}{\Phi_0}\right)^{1-\alpha}\bigg|^{\Phi_2}_{\Phi1}\,,$$

\noindent
where $n_0 = 3.14 \times 10^{-14} \; \textrm{cm}^{-2}\; \textrm{day}^{-1}$ and
$\Phi_0 = 10^{16}$ Mx. Remembering that they found a power index $\alpha =
-2.7$, we obtain a frequency of $N_{\rm ev} = 4.9 \times 10^{-17}\;
\textrm{cm}^{-2}\; \textrm{day}^{-1}$ for a flux range from $5.5\,\times 10^{17}$ 
to $1\, \times 10^{20}$ Mx. Taking into account our FoV of $\simeq
1300\;\textrm{Mm}^{2}$ and the duration of the observations of 30 minutes,
we get $N_{\rm ev} \simeq 14$ events (i.e., 7 bipoles). The observed number of
flux patches within this range in our data set is smaller than the prediction
from the \citet{Thornton:11} work, as we found only the two footpoints of the
bipole. Note also that no flux patches with fluxes above $6.5 \times 10^{17}$
Mx are observed, whereas the above distribution would have suggested otherwise.
Given the short duration of the period analysed here, it is unclear whether
this disagreement is fortuitous or indicates an excess in the predicted number
of bipoles due to an overestimate of the total density of emergence events.
The activity minimum in which the Sun was residing during the \textsc{Sunrise}
flight could also be behind our low number of observed bipoles.
Note that the difference in both time and space resolution between the 
two data sets could affect this comparison: thus, it would be worthwhile applying 
the algorithm used by \citet{Thornton:11} to these \textsc{Sunrise} observations.

On the other hand, the bipole closely behaves as a small-scale version of an
ER: we found a good agreement with the dependence of the footpoint separation
with the flux content proposed by \citet{Hag:01}. The overall morphology
is quite similar, even if we do not find any evidence of mixed polarities
in between the two footpoints during the emergence, as reported by 
\citet{Guglielmino:10} for an ER with flux content $\sim 1.5 \times 10^{19}$ 
Mx, i.e., twenty times larger than the bipole studied here. 

Further studies of similar time series can be of great value to consolidate
the results presented here. More cases are needed to clarify these findings, using 
spectropolarimetric data of a similar resolution, sensitivity and temporal coverage
such as those that could be obtained with specially tailored observations with 
SOT/\emph{Hinode}, CRISP/SST, and IBIS/DST. We are not aware of such results yet and,
thus, it is difficult for the moment to extend and compare these findings with more 
examples, and to understand their implications for the global solar flux budget.

\acknowledgments

This work has been partially funded by the Spanish Ministerio de Educaci\'on y
Ciencia, through Projects ESP2006-13030-C06-01/02/03/04 and
AYA2009-14105-C06, and Junta de Andaluc\'ia, through Project
P07-TEP-2687, including a percentage from European FEDER funds. The German
contribution to \textsc{Sunrise} is funded by the Bundesministerium f\"ur Wirtschaft und
Technologie through Deutsches Zentrum f\"ur Luft- und Raumfahrt e.V. (DLR),
Grant No. 50 OU 0401, and by the Innovationsfond of the President of the Max
Planck Society (MPG). Financial support by the European Commission through the
SOLAIRE Network (MTRN-CT-2006-035484) is gratefully acknowledged. S.L.G. and L.R.B.R. 
acknowledge insightful discussions on small-scale flux emergence processes 
within the ISSI International Team lead by K. Galsgaard and F. Zuccarello 
at the ISSI (International Space Science Institute, Bern), and the ISSI for support. 
This work has been partially supported by the WCU grant (No. R31-10016)
funded by the Korean Ministry of Education, Science and Technology. 
Use of NASA's Astrophysical Data System is gratefully acknowledged.

{\it Facilities:} \facility{Sunrise}.

\end{document}